\documentclass[sigconf,authorversion,natbib=false]{acmart}
\usepackage{graphicx} 
\usepackage{url}
\usepackage{epsfig}
\usepackage[utf8]{inputenc}
\usepackage[url=false,isbn=false,eprint=false]{biblatex}
\title{Steps towards an Ecology for the Internet}
\subtitle{Computing through crisis into the next decade of growth}

\author{Anil Madhavapeddy}
\authornote{Both authors contributed equally to the paper}
\affiliation{
  \institution{University of Cambridge}
  \country{}
}
\orcid{0000-0001-8954-2428}

\author{Sam Reynolds}
\authornotemark[1]
\affiliation{
  \institution{University of Cambridge}
  \country{}
}

\author{Alec P. Christie}
\affiliation{
  \institution{Imperial College London}
  \country{}
}

\author{David A. Coomes}
\affiliation{
  \institution{University of Cambridge}
  \country{}
}

\author{Michael W. Dales}
\affiliation{
  \institution{University of Cambridge}
  \country{}
}

\author{Patrick Ferris}
\affiliation{
  \institution{University of Cambridge}
  \country{}
}

\author{Ryan Gibb}
\affiliation{
  \institution{University of Cambridge}
  \country{}
}

\author{Hamed Haddadi}
\affiliation{
  \institution{Imperial College London}
  \country{}
}

\author{Sadiq Jaffer}
\affiliation{
  \institution{University of Cambridge}
  \country{}
}

\author{Josh Millar}
\affiliation{
  \institution{Imperial College London}
  \country{}
}

\author{Cyrus Omar}
\affiliation{
  \institution{University of Michigan}
  \country{}
}

\author{William J. Sutherland}
\affiliation{
  \institution{University of Cambridge}
  \country{}
}

\author{Jon Crowcroft}
\affiliation{
  \institution{University of Cambridge}
  \country{}
}

\copyrightyear{2025}
\acmYear{2025}
\setcopyright{rightsretained}
\acmConference[AAR 2025]{The sixth decennial Aarhus conference: Computing X Crisis}{August 18--22, 2025}{Aarhus N, Denmark}
\acmBooktitle{The sixth decennial Aarhus conference: Computing X Crisis (AAR 2025), August 18--22, 2025, Aarhus N, Denmark}
\acmPrice{}
\acmDOI{10.1145/3744169.3744180}
\acmISBN{979-8-4007-2003-1/25/08}

\begin{document}

\begin{abstract}
The Internet has grown from a humble set of protocols for end-to-end connectivity into a critical global system with no builtin ``immune system''.
In the next decade the Internet will likely grow to a trillion nodes and need protection from threats ranging from floods of fake generative data to AI-driven malware.
Unfortunately, growing centralisation has lead to the breakdown of mutualism across the network, with surveillance capitalism now the dominant business model.
We take lessons from from biological systems towards evolving a more resilient Internet that can integrate adaptation mechanisms into its fabric.
We also contribute ideas for how the Internet might incorporate digital immune systems, including how software stacks might mutate to encourage more architectural diversity. We strongly advocate for the Internet to ``re-decentralise'' towards incentivising more mutualistic forms of communication.
\end{abstract}



\maketitle

\section{Introduction}

Ecology is the study of relationships between living things and their environment across all scales, from blue whales in the ocean, to bacteria living on the surface of a leaf. It encompasses all life processes, interactions and adaptations; community structure and composition; competition and foodwebs; ecosystem structure and energy flow; landscapes and life histories; mutualism, parasitism and infectious disease; the abundance and distribution of organisms; and how ecosystems function~\cite{wiki:Ecology}.
Meanwhile, the rather more abiotic Internet is the global system of computer networks that standardises protocols to communicate among each other. It lacks any central control structure and is a network of networks that connect private, public, academic, business and country networks, linked by a broad array of physical connection technologies~\cite{wiki:Internet}.

Bateson's classic work ``Ecology of Mind'' notes that {\em ``the history of evolutionary theory is inevitably a metalogue between man and nature, in which the creation and interaction of ideas must necessarily exemplify evolutionary process}''~\cite{Bateson2000-pg}.  In this paper, we explore the idea of an ecology for the Internet, and how the Internet in its current phase of growth can draw parallels from ecological theory to become more resilient, sustainable, and trustworthy into the coming decades as it approaches a trillion connected nodes~\cite{armtrillion}. To help guide our discussion, we will first clarify our teleology. We are using the term {\it ecology} directly, not merely as a weak metaphor for a collection of stakeholders in a socio-techno-economic system.
We examine those dimensions of ecology that have no intentional purpose~\cite{Knight2021} and serve to preserve and advance life through selection pressures against competing strategies~\cite{bar2005biomimetics}. The Internet, now entering its fifth decade of existence~\cite{paloque2019arpanet}, needs to continue to be based on system designs that provide communication, whatever the adversary does; whatever and whoever the adversary is, even if the adversary is  sometimes ourselves. 

\subsection{The Internet as an Ecology}
\label{s:internetecology}

The Internet started from humble beginnings~\cite{coffman1998size} and has now grown into a behemoth of interconnected nodes with zettabytes of traffic flowing across it daily~\cite{ciscoreport}. As growth continues to multiply, it is facing new existential threats; from the rise of nation-state actors, to the increasing sophistication of cyber-criminals, and the increasing complexity and layering of its protocols  resulting in the ossification of its infrastructure~\cite{handley2006internet}. Accelerating all of these threats are emerging AI technologies that can automate attacks in superhuman timescales~\cite{arxiv.2404.08144}, poison the wells of human information with fake or low-quality generated data~\cite{Runco2024}, and even manipulate the physical world through Internet of Things (IoT) devices~\cite{10.1145/3355369.3355577}. 

\begin{table*}[h]
    \centering
    \begin{tabular}{|l|p{9cm}|p{1cm}|}
    \hline
    \textbf{Biological Concept} & \textbf{Internet Concept} & \textbf{Section}\\
    \hline
    DNA & Protocol specifications (RFCs), file formats & \S\ref{s:mutatis} \\
    Genes & Software modules, protocol functions, reusable code &  \S\ref{s:resilience} \\
    Proteins & Operating system processes, code regions & \S\ref{s:containment} \\
    Virus & Malware, botnets, Internet worms & \S\ref{s:antibotty} \\
    \hline
    Chromosome & Software interfaces, formal specifications & \S\ref{s:resilience}\\
    Organelle & CPUs, RAM, NICs, SSDs & \S\ref{s:containment} \\
    Cell & End nodes (IoT devices, servers, middleboxes) & \S\ref{s:mutatis} \\
    Tissues & Clusters of hosts (data centers, edge clusters) & \S\ref{s:scale} \\
    Organs & Domain-specific networks (e.g., CDN nodes) & \S\ref{s:scale} \\
    \hline
    Organisms & Entire Internet services (e.g., Google, Facebook) & \S\ref{s:trust} \\
    Populations & Device ecosystems, software monocultures & \S\ref{s:trust} \\
    Communities & Federated networks (e.g., ActivityPub, Bluesky) & \S\ref{s:proof} \\
    Ecosystems &  Overall Internet resilience & \S\ref{s:proof}\\
    \hline
    \end{tabular}
    \caption{Just as biological processes exhibit self-organization, we explore in this paper how these principles might apply to the network of networks that is the Internet. The concepts are in rough order of increasing scale.}\label{tab:bio-internet}
\end{table*}

The core purpose of the Internet is to facilitate communications in a decentralized, scalable and extensible manner. It is based on the end-to-end principle~\cite{Kempf2004} that dictates the network itself remain as simple as possible, with complexity pushed to the software on end hosts instead. This has permitted decades of evolutions of communications protocols ranging from the original ARPANET to the modern incarnation~\cite{paloque2019arpanet}. However, the end-to-end principle has also led to the Internet's current vulnerability to automated attacks as the network itself has no builtin ``immune system'' to protect itself with, and end hosts cannot upgrade quickly enough.

However, individual Internet hosts can also deploy new software techniques in small clusters without requiring central ``permission''.  In this sense, they are very similar to the phenomenon of self-organisation that is primarily seen in biological systems. We therefore turn to ecological inspirations, such as plant-pathogen interactions and evolutionary theory, to help us to build a more resilient Internet with more active defence systems.

Table~\ref{tab:bio-internet} contains an illustrative mapping between biological and Internet systems, beginning firstly from the smallest components and then scaling up to entire populations. Throughout the rest of this paper, we will explore how these biological concepts can be applied to the Internet. Central to our discussion will be we can learn from the natural world to embed new concepts into the Internet architecture that will enable evolution towards a more resilient system despite the number of nodes growing into the trillions. We will also sketch out how some these ideas could be implemented in practice, and what challenges we may face in doing so. Our primary goal is to inspire further research and discussion on this topic, rather than to provide a comprehensive solution.

\section{Resilience against attack}
\label{s:resilience}

\begin{figure}
  \includegraphics[width=\linewidth]{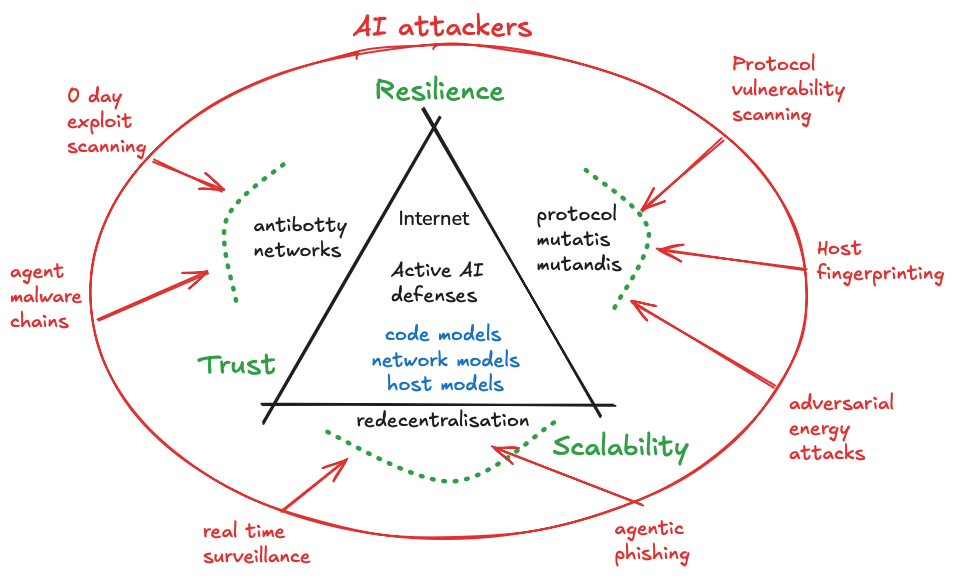} 
  \caption{The Internet needs to urgently evolve active defenses against new threats from AI-powered attacks. We discuss three such defences in this paper.}\label{fig:arch}
  \end{figure}

The Internet protocol culture is one of open standards and collaborative processes. Rather than being dictated by central authorities, Internet standards emerge through ``rough consensus and running code'', meaning ideas are adopted when working implementations demonstrate real-world value. Design principles like Postel's Law (\textit{``be liberal in what you accept, conservative in what you send''}) have encouraged resilience and compatibility~\cite{Kempf2004}, while the end-to-end principle keeps the core network simple and pushes complexity to the edges~\cite{McQuistin2021}.
In theory, this should lead to a diverse ecosystem of implementations by which newer software revisions can extend the protocols in a backwards compatible manner to older endpoints. However, in practice, only a few pieces of software dominate most endpoints. Consequently, if a vulnerability is discovered in these dominant implementations, millions of hosts become immediately susceptible to compromise. And because of the low latency interconnectedness of the vast majority of Internet hosts, this compromise can and has happened within minutes~\cite{staniford2002own}.

We can view this widespread software monoculture as being similar to an evolutionary bottleneck: working towards the currently-optimal Internet structure took place under a certain set of tight and idealistic environmental constraints of the time (e.g.~hardware and technology availability, an early push for openess and simplicity), which left the network as the fittest version of itself at the time, best adapted to the environment it faced. However, the resultant lack of diversity is now a threat as the pressures and environmental conditions change. The Internet now currently lacks the flexibility to respond to challenges, and these are weaknesses consistent across the entire population due to the software monoculture.  The most obviously visible threat is the rise of botnets, which can commandeer vast numbers of machines to launch further attacks~\cite{Andriesse2015}. To counteract this, a more resilient model would promote active diversity in the software stack~\cite{Larsen2014}. This model would not only fight against vulnerabilities but also provide shielding to hosts that cannot deploy such countermeasures (consider, for instance, a pacemaker that provides a critical service but cannot be easily updated). A more resilient network would ensure that such devices are protected from potential threats via cooperation and local incentives. Ribiero et al argue that a successful approach to evolving the Internet in such a manner requires backwards compatibility, an evolutionary strategy from the current Internet, and room for new architectures to proliferate~\cite{ribeiro2024internet}. This is a good set of principles to start from; given the rapid growth of botnet- and AI-assisted attacks, we need to build more active translation layers that can adapt (ideally unpredictably) in real-time.

Biological ecosystems also have self-regulation mechanisms~\cite{Lenton2002} where the population of a species is controlled by the availability of resources and nutrient cycles. 
We know that diverse types of microbes within microbial ecosystems can specialise in distinct functions, contributing substantially to the overall stability of the ecosystem~\cite{Kost2023}. This also allows for instances of cooperation whereby microbes form alliances\footnote{Lichens are a model example of alliances between fungi and algae or cyanobacteria~\cite{allen2022call}.} for resource acquisition~\cite{Mostafa2024}, analogous to how Internet routing is a set of cooperative peering agreements~\cite{norton2011internet}. 
This cooperative imperative also extends across inter- and intra-species relations~\cite{Bronstein2024}.
In contrast, Internet-connected hosts tend to require individual active management in order to stay connected and secure. Within a network, the collection of hosts is not controlled by the availability of resources, and malware-laden machines can lie dormant and be controlled by botnets for years.

In our first thought experiment, we will discuss an approach to self-regulating Internet networks that can adapt to changes in their environment and take active measures to control the network.

\subsection{Active Antibotty Defences}\label{s:antibotty}

Consider the example of detecting and stopping malware within an urban community. In such a deployment, there are relatively small clusters of hosts that have multiple network connections to each other; for instance, a mobile phone host might gain access to all hosts within another home network when the owner visits a friend and connects to their wifi~\cite{Piasecki2021}.  The current process of protecting hosts is laboriously manual, involving installation of malware detectors on every endpoint and ensuring the antivirus systems are themselves up-to-date and not themselves going to become routes of exploit.  Biological systems operate their defence mechanisms much more autonomously though, via immune systems which fight back routine disease incursions without affecting the overall biological host adversely in routine infections~\cite{parkin2001overview}. If we consider malware in the same light as a biological virus, then what is the equivalent of antibodies to automatically detect and eliminate malware in a friend's home who may not be actively maintaining software updates on all their devices?

One approach would be for every host to act as an antibody\footnote{Or more precisely, act as a T-cell or B-cell; the cells of the immune system which seek out antigens and eliminate them with antibodies or other methods.} and actively scan the network for signs of malware on nearby friendly hosts. This immediately has a temporal advantage over botnets that must scan whole netblocks~\cite{antonakakis2017understanding} to find vulnerable hosts, since each host only checks its local population. When a local host does find evidence of malware, it could take active measures to either isolate the infected host from the network, or directly exploit it before a global botnet does to deliberately patch it. Since the local host is acting in the interests of the immediate community and advertising its intent, it is acting as a vigilante~\cite{10.1145/1095809.1095824} to protect the community network from harm, and is not incentivised to exploit the host for personal gain.

Such an antibotty\footnote{``Antibotty'' is our portmanteau of ``antibody'' and ``botnet''} community network becomes stronger as the community hosts increase their diversity of operating systems and application services (\S\ref{s:mutatis}) since the chances of one of them spotting a vulnerability in their peers increases. There is also the possibility of building higher level services to repair the damage caused by isolating a host; nearby hosts might set up application-level proxies to reroute traffic and even filter traffic until the vulnerable host can be patched.  In an extreme case where no patch exists, the offending host could be ``killed'' by terminating its network connection from its peers --- this might happen for an old IoT device which has not received software updates due to obsolescence or repeated takeovers~\cite{Piasecki2021}. The antibotty network approach outraces global botnets because local hosts only need to scan orders of magnitude fewer devices in their vicinity, and could therefore be a step towards building a self-regulating Internet ecology. It also turns a weakness into an advantage by using the sophisticated command-and-control mechanisms that have been developed by global botnets for malicious use and deploying them within the community~\cite{StoneGross2009}.

There is a reason that antibotties are currently a thought experiment though, as great care needs to be taken to ensure that the antibotties do not themselves become a threat. Since every host acts as a vigilante, a structure is needed to ensure that the antibotties do not cause unintended harms. An individual node cannot easily distinguish between mission-critical devices (e.g.~a pacemaker) and a non-critical device (e.g.~a smart lightbulb in a hallway).
Biological systems deal with this via regimented, systematic and ordered processes of development to form a hierarchy of capabilities and protections. This is why, for example, an organism's heart follows an exact blueprint, or legs do not appear where antennae should be on a fruit fly.\footnote{Assuming healthy development and the absence of homeotic transformations.} As part of this process, immune systems develop which have multiple ways of responding to external threats (e.g. from mucus to skin), through which different sorts of white blood cells that combat a range of antigens. An important part of this immune system is the blood-brain barrier, which is a semi-permeable membrane that is highly selective about what can pass across to the vital brain from the rest of the body, and antibodies typically cannot because they would cause harm to delicate functions~\cite{abbott2010structure}.

A similar design for the Internet version of immune systems would be to have a hierarchy of controllers to coordinate clusters of antibotties. The design needs to be extended into the network protocols by which vigilante hosts communicate, and the network structure needs to remain resilient to attacks that might exploit the vigilante mechanisms themselves. Like biological systems, antibotty networks should establish multiple layers of scoped protections against threats, for example via opportunistic proxies and neighbourly packet inspection on local broadcasts. Although Internet Service Providers have not historically been keen on active countermeasures~\cite{10.1145/1198255.1198265}, this attitude may need to change to adapt to the rising tide of AI-powered attacks.

The communication mechanisms behind antibotty alerts amongst each other also need some care to not leak private metadata. With existing vigilante mechanisms~\cite{10.1145/1095809.1095824} the alerts also advertise the mechanism behind the attack, which could be exploited by a new adversary to join the network and learn new tricks. One solution would be to seal the alerts and only allow trusted nodes to decode them, but a more radical ecologically inspired mechanism would be to introduce more mutations throughout the software stack in response to the new vulnerability. The biological antibodies produced by species, are by design, incredibly diverse via a process called ``somatic hypermutation''~\cite{Ridani2024} whereby mutations are introduced within B-cells to create a diverse range of antibodies. They have error-prone DNA repair and copying proteins by design to create the highest diversity of antibodies possible. We consider how to emulate this in digital antibotties next (\S\ref{s:mutatis}).

\section{Scaling the Internet towards a trillion nodes}
\label{s:scale}

As the Internet inexorably grows, it needs to do so in a manner that does not make it more brittle to global disasters --- either from anthropogenic ones such as malware as discussed earlier, or natural occurrences~\cite{10.1145/3452296.3472916}.
The Internet has been quite successful at providing diversity in the physical layer, and transmission systems can use anything that can convey sufficient signal to get an IP packet almost anywhere in the global via RF, optical, electronic, or even mechanical routes. From the IP layer up, we are largely in the domain of software systems which are much easier to reconfigure and program. The key constraint is that we still have to interoperate via protocol specifications that we collectively agree on. As long as network endpoints can understand each other, there is no reason not to have a wide range of such protocols --- after all, humans have a wide range of languages, and the natural ecosystem has a much wider range of ways to convey signals between members of same and different species~\cite{Sharma2024,arxiv.2502.00344,wang-etal-2024-phonetic} and then even within and between the cells of those individual species..

Mutualistic relationships built over these signals abound between creatures in the wild; in the African Savanna creatures who can smell, hear or see better than each other such as warthogs, impala and giraffes~\cite{Leuthold1977} cooperate to give warning to each other of possible predators~\cite{Diplock2018}. In a natural ecosystem ``indirect benefits play the main role in explaining cooperation within species''~\cite{West2021}, giving us encouraging signs that we need not advocate for purely immediate metrics of cooperation. Increasing host endpoint diversity would not reduce cooperation as long as there remains consensus on the protocols that allow them to communicate, and indeed in some areas such as wireless networks cooperation is necessary for effective use of the frequency spectrum~\cite{1362898}.

How far could we take this diversity within Internet nodes? There exists a staggering diversity of species in nature; mostly millions of varieties of insects~\cite{Stork2018} and fungal species~\cite{Purvis2000} which in turn influence the development of plants and other aboveground species~\cite{vanderHeijden1998}.
These naturally biodiverse systems have evolved to have extreme resilience to changes in the environment, whether from global changes caused by climate changes, pollution, or coincident anthropogenic conversion~\cite{Thorogood2023}. And conversely, as ecosystem biodiversity reduces in the anthropocene, the resilience of these natural systems is also declining rapidly~\cite{Oliver2015}. Monocultures in nature not only lead to ecosystem-wide brittleness (such as wildfires having outsize impact on the broader population~\cite{wildfire2017}), but they also close off whole strands of evolutionary exploration. The same holds in electronic systems where a single vulnerability can lead to global compromise~\cite{moore2003inside}, thus leading to our next thought experiment.  

\subsection{Protocol Mutatis Mutandis}
\label{s:mutatis}

A dangerous monoculture of software implementations for end nodes has arisen on the Internet. A few operating systems, namely Windows, macOS and Linux, dominate the landscape, and certain aspects, particularly network stacks, tend to evolve slowly. This ossification occurs for several reasons: the difficulty of changing the userspace APIs that applications use to access the filesystem and network stack~\cite{rise_of_libos}, the deployment of middleboxes in the network which adopt a fixed and often overly strict interpretation of protocols~\cite{raiciu2012hard}, and (perhaps a controversial position to take) the huge success of the open-source model which encourages code reuse~\cite{Feitosa2020}.

This all implies that the Internet's natural state is heading towards one of ossification and centralisation from a software perspective, despite the diversity in physical link mechanisms. What we need is a mechanism that encourages ``mutations'' to develop in the software stacks of Internet hosts, and a way to measure the effects of these mutations on the network as a whole. This is a protocol implementation evolution that is not driven by a central authority, but by the individual hosts themselves, while remaining respectful of the communication specifications themselves to remain understandable by their peer hosts (which might themselves also be mutating).

The open-source process of code reuse and sharing between different software is similar to Horizontal Gene Transfer~\cite{Arnold2021} whereby, non-sexually, bacteria and archaea and eukaryotes can release DNA (genes) into the environment and be taken up by other cells which can incorporate them directly into their genome (codebase). It has contributed to antibiotic resistance in bacteria, the evolution of eukaryotic cells (all animal cells are eukaryotic) and their endosymbiosis with mitochondria. It allows genetic information to pass between species that would be unable to reproduce sexually as they are too far removed from each other evolutionarily. Doyle observed that the Internet IP layer possesses a similar ``deconstraining constraint'' for electronic communications; IP has a specification that still permits a wide range of interpretations over it~\cite{doi:10.1073/pnas.1103557108}.

A key aspect of such mutations in software is that they would not just be driven by their genes (that is, the software that was originally written for that host), but could also be guided by the environment within which the hosts are deployed. The Internet, after all, spans everything from remote terrestrial sensors to satellites to embedded pacemakers to vast datacenters. The conventional view of evolutionary adaptations is that they arise from the natural selection of random DNA mutations across generations~\cite{Jablonka2025}, which would be far too slow for the Internet.  However, there has been a decade-long debate in evolutionary biology about a different approach dubbed ``extended evolutionary synthesis''~\cite{Laland2014} that argues that development driven by dynamic conditions also drives how an organism behaves~\cite{Lala2024-ue}. They note that ``evolution must proceed where development leads'', which is a useful perspective for our desire to broaden the Internet's software diversity.

\begin{figure}
    \includegraphics[width=\linewidth]{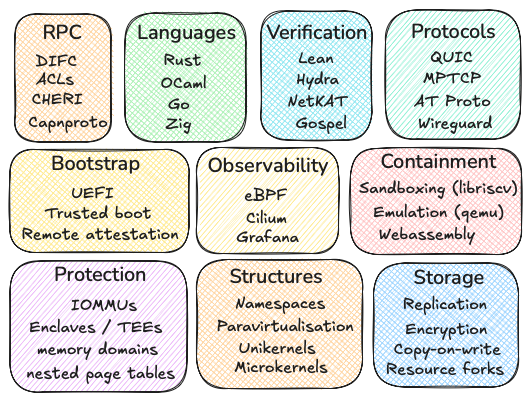} 
    \caption{The modern software stack has a number of points where automated patch mutations can be inserted, contained, evaluated and stored without affecting the whole system.}\label{fig:stack}
    \end{figure}
            
One emerging mechanism available to kickstart evolution in modern software systems is the rise of AI-driven code models, which are increasingly proficient at program synthesis~\cite{Li2022}. While these do not yet generate code as well as a human author~\cite{NegriRibalta2024}, with sufficient guardrails, it is possible to create patches for existing code.
By developing a system that can patch endpoints incrementally and measure the effects across the entire network, we could perform directed evolution~\cite{Tian2024} within the software stack more rapidly.
Any AI-driven code generation needs adequate context in order to satisfy the constraints of such code mutation. The Internet community has long maintained~\cite{Khare2022} a rich corpus of plaintext ``RFC'' specifications that can be interpreted via Large Language Models into formal specifications~\cite{10.1145/3626111.3628205}, which can guide LLMs towards generating only permissible protocol messages in code mutations.

\subsection{Containment of Mutations}
\label{s:containment}
Much like the blood-brain barrier regulates the transfer of chemicals between the circulatory and nervous systems in a biological system, operating systems must also regulate the impacts of software mutations. There are  now numerous hardware and software containment layers where code mutations can be inserted safely and with minimal overhead. Figure~\ref{fig:stack} shows some possible hooks:

\begin{itemize}
\item Hardware protections, such as IOMMUs, secure enclaves, nested paging, capabilities and intra-process protections all allow for code mutations to be guarded.
\item New operating system structures, such as unikernels and process namespaces, as well as observability mechanisms such as eBPF allow for dynamic checks to be inserted across the software layers.
\item Lightweight sandboxing mechanisms, such as Webassembly, are widely deployed in web browsers and cloud environments, allowing for experimental code to be guarded and distributed easily.
\item Storage systems now extend traditional POSIX storage interfaces~\cite{2015_sosp_sibylfs} to include support for copy-on-write snapshots, resource forks, encryption and incremental replication, allowing for reliable storage in the face of mutations.
\item Newer protocols such as QUIC, Wireguard and the ATProto~\cite{Kleppmann2024} are all implemented in userspace, allowing for much easier modifications.
\item Formal verification frameworks such as Hydra~\cite{10.1145/3603269.3604856} and NetKAT~\cite{10.1145/2535838.2535862} can be used to verify the effects of changes on the network before they are deployed, providing a safety net to routing health.
\item Systems languages such as Rust, Go, OCaml and Zig continue to rise in popularity, providing layers of type safety to code changes proposed, as well as formal verification tools such as F* and Lean that provide stronger guarantees for the outputs of code models.
\end{itemize}

There remain significant challenges to overcome; since malware has free reign to exploit weaknesses, there is a risk that adversarial attacks~\cite{10.1145/3688839} could allow malware to guide code mutations down undesirable paths that open up new exploits.
The same is exactly true in biology as well; viruses hijack cell machinery to make the proteins they need to build more viruses and mimic (or crack) the keys other proteins use to enter cells and hijack them. Research in cell biology has shown that ``protein language models'' can evolve human antibodies by suggesting mutations that are evolutionarily plausible~\cite{Hie2023}. The same approach could generate code mutations that are protocol compatible for antibotty hosts.

A safer and simpler approach would be to use capability-based programming and hardware extensions~\cite{watson2015cheri}. In this model, a piece of code uses a piece of data as both the key and the lock to access a mutual resource. This would allow for both fine-grained tracking of the sources of mutations, and also to craft more sophisticated security policies for individual layers in the software stack~\cite{2024_hope_bastion}.

There is also a higher level question about how extensive such patches should be. The existence of sexual reproduction is a puzzle to evolutionary biologists~\cite{Otto2009}, as with each generation females lose half of their genetic material (albeit it is duplicated) and in many cases have no parental care in exchange!  The most likely reason for this is that sexual reproduction results in rapid new combinations of material. These combinations may result in produced diverse lifeforms that are not only exciting new options but are also are different from each other and the parent, so a pathogen or predator that has adapted to the parent faces new genotypes. This is especially important where the pathogens have shorter generation times than the host such as a long-lived oak tree exposed to environmental stresses~\cite{wargo1996consequences}.  So when considering the extent of mutation introduced as part of our software mutatis, we might also swap out entire subsystems with alternative implementations as form of ``sexual component mixing''.  There are already examples of this within components such as TCP ex machina~\cite{10.1145/2486001.2486020}, and across operating system components via unikernels~\cite{2013_asplos_mirage} that are making their way into conventional operating systems like Linux~\cite{10.1145/3552326.3587458}.

\section{Mutualism and Trust}
\label{s:trust}

Trust in biological networks can be established because there is constant meaningful and consequential feedback between all parties involved, ranging from the smallest cells to the largest mammals.
Many of these relationships are symbiotic in structure to both each other and their context~\cite{Chomicki2022,DelDottore2023}. At a macro level, bird breeding populations, for example, are affected by their local environment network~\cite{Lin2023}. At a micro level, social interactions among viruses occur whenever multiple viral genomes infect the same cells, hosts, or populations of hosts~\cite{Leeks2023}.

One of the most remarkable networks on the planet is the underground mycorrhizal fungal system, which are the dominant form of nutrient exchange between land plants and the soil~\cite{wohlleben2016hidden}. The earliest land plants were mycorrhizal\footnote{The Rhynie Chert, a 407-million-year-old fossilized geothermal site, provides crucial evidence of early land plant-fungal associations, including mycorrhizae.} and the root networks that we normally associate with plants only evolved 400 million years after the first fungal networks, as an evolutionary optimisation to facilitate the symbiotic relationship~\cite{StrulluDerrien2018}.
 
Mycorrhizal networks exist because the fungi take sugars from plants ``in exchange'' for moisture and nutrients gathered from the soil by the fungal strands. If trees stopped sharing, it would not be long before the fungi did too.
These fungal networks have evolved further to longer range facilitate ``nutrient trading networks'' that respond to resource inequality by withholding or supplying them to distant nodes~\cite{Whiteside2019}. Individual fungal nodes have specialised, with subsets of polymorphic nuclei acting cooperatively, or as traders, or entirely selfishly~\cite{No2018}. The mycorrhizal network itself acts as a key driver for aboveground plant population and biology~\cite{Tedersoo2020}, which in turn provides habitats for many above-ground species.
A key point here is that the soil is packed with pathogenic fungi, so plants and fungi have developed intricate handshakes which set the scene for this extraordinary collaboration that happens trillions of times a day~\cite{Bonfante2010}, but the protocol has yet not been exploited by competing species to ``steal'' the nutrients!

The Internet looks remarkably similar in many ways, with individual endpoints specialising to certain tasks, and protocols existing for peer-to-peer cooperation. There are longer-range consensus protocols to symbiotically maintain the health of the larger network, such as BGP and DNS that depend on peer communication via a hierarchy. And the Internet substrate itself provides a ``habitat'' for the billions of humans that use it every day to communicate among each other. The Internet has never mandated a specific identity model beyond providing addressing and naming mechanisms, instead leaving this to hosts to negotiate. This has led to a healthy proliferation of identity architectures across layers of the protocol stack, from simple username and password combinations, to public key infrastructure, and biometric passkeys.  While the early Internet, with its smaller population, exhibited strong mutualism and cooperation, we are now seeing a rise in more parasitic models that break down trust across the network.

The first threat is the rise of advertising driven business models that lead to a centralised attention economy~\cite{menczer2020attention}. Trackers deployed across the Internet by a few large providers have lead to extremely high resolution user fingerprinting and data collection mechanisms~\cite{Bujlow2017}. Google, for example, uses a vast array of third party trackers as ``secure signal providers'' who share secure signals with advertising bidders competing from clicks from humans.\footnote{\url{https://support.google.com/admanager/answer/10488752?hl=en}}

The second threat is that even such extreme tracking mechanisms are themselves under pressure, as they are not scalable to billions of users who need to prove their ``humanity'' for websites trying to distinguish between real people and fake clicks~\cite{dave2012measuring}.
AI has recently made it vastly easier to impersonate humans, and to create fake data that can be used to manipulate these advertising models. This is leading to a crisis of trust of information accessed over the Internet, where it is difficult to determine whether both a given piece of data or the creators are genuine or not. The problem is particularly acute in the context of social communications, where fake news can spread rapidly and cause real-world harm~\cite{lazer2018science}. The move online has changed the economics of news distribution to one of ``surveillance capitalism'', contributing to the lowering of diversity of news sources~\cite{10.1145/3292522.3326019} and leaving just a few big players standing who serve a large population.

Mutualism in biological networks --- both long-term symbioses and brief exchanges --- have transformed entire ecosystems and operate alongside intense competitive interactions for resources~\cite{Bascompte2008}. Humans, on the other hand, don't currently have the ability to easily fight back against the attention-sapping concentration of media arising from both centralised advertising models that are augmented by generative artificial intelligence.  The good news for the Internet is that, ultimately, species that ``cheat'' (that is, undermine the cooperative social contributions of others to further themselves) are often punished by the group and suffer because of it. These cheaters also tend to become less fit over time than other species and cheat out of necessity to maintain their status.  Cooperation evolves as species choose cooperative partners, which begets more cooperation~\cite{Riehl2016}.
If we consider the current structure to be parasitic in nature, where a few large corporations benefit at the expenses of the larger population of billions, then the Internet is approaching a tipping point where the billions spent on advertising to capture trillions of hours of humans' attention will be flooded by fakery and will collapse. Ecology tells us that a likely --- but not guaranteed~\cite{SACHS2006} --- stable fit state after we recover from the parasitic assault is one of mutualism, where the network itself can manage cooperation locally.

\subsection{Redecentralisation}
\label{s:proof}

A decade ago in Aarhus, we made an argument for ``databoxes''~\cite{2015_aarhus_databox}, an architecture that shifts private data processing to run locally rather than centrally for sensitive information such as health metrics~\cite{zhao2020privacy} or travel~\cite{2012_mpm_caware}. Databoxes never took off in the years since, but --- as with the unpredictable pace of evolutionary selection --- the time may now be right for a redecentralisation of the Internet to emerge!
Most Internet trust models have pooled around a monoculture of identity providers who control the roots of trust for almost all of our online communications, with anonymity relegated to specific usecases~\cite{2010_iswp_dustclouds}. But some systems such as DNS and e-mail retain (limited) decentralisation~\cite{10217834}, and the fabric of the Internet still allows for plenty of innovation due to the end-to-end principle allowing for new protocols to emerge.

Fake news is of weak quality (or of ``lower fitness'') than real news, but its scale and our inability to organise widespread community pushback against it, makes it difficult to punish and allows it to proliferate. A new breed of social network architectures are now emerging to provide that pushback. ActivityPub is one such protocol standardized in the W3C that permits decentralised hosts to communicate with each other for a variety of services such as messaging, microblogging, and photo and video sharing.  Bluesky is another, with an underlying ``ATProto''~\cite{Kleppmann2024} that has grown to over 30 million users by Feb 2025 and supports having multiple interoperable providers for every part of the system. One consequence of decentralised algorithms in these systems is that they fundamentally affect how users communicate and perceive each other's communications with greater agency as to where to focus their attention~\cite{surveshamrajmehta2024}.

It has also been getting simpler for users to {\em deploy} their own infrastructure. Serverless cloud providers can execute custom user code, custom DNS infrastructure allows for users to own their own domain names and identity online~\cite{2013_foci_signposts}, and a self-hosted personal clouds\footnote{For example, \url{https://github.com/RyanGibb/eilean-nix}} permit the use of open standards and federated protocols for individual groups and families and organisations to communicate with each other.  What's missing is the symbiotic infrastructure to protect and nurture such activities, which we hope can arrive via mutualistic new protocols that we have discussed here such as antibotties (\S\ref{s:antibotty}) and protocol mutatis mutandis (\S\ref{s:mutatis}). Mutualism in biological networks is also not necessarily the only stable endpoint, with reversion to autonomy also being regularly observed~\cite{SACHS2006}, so we should not be surprised if the Internet also oscillates between centralised and decentralised models over the coming decades.

\section{Caveats}\label{s:caveats}

There are several aspects of ecology which we have not discussed in this paper with relation to the Internet's growing pains. Most obviously, mutualism only works for some biological processes but not for others --- it is ideal for nutrient exchange, but not when competing for mates!
One aspect of the natural world that we have not included in this discussion, but merits detailed future attention, is {\em expiry}.
In Venki Ramakrishnan's work on ``why we die''~\cite{ramakrishnan2024we} he makes it abundantly clear that there are as many reasons why individuals and systems (populations) arise and fall. 
In nature, pathogens cause the death of individual cells and bodies, but the fittest survive. In the future we intend to consider how the Internet should incorporate this; we have dealt with this for e-mail spam, for example, but the result has been an increasingly centralised monoculture~\cite{10217834}.

Ecology also teaches us that the local spread of disease can be contained by the evolution of disease response, but global rapid spread of diseases (from Covid 19 to ash tree dieback~\cite{pautasso2013european}) can be catastrophically rapid.  Local defence systems developed over millennia to cope with local diseases do not have the information they need to cope with newcomers, and so some restriction of the flow of information across Internet nodes within political boundaries might actually be critical to ensure the long-term health of the clusters within the larger ecosystem.

It is not yet clear whether the evolution of Internet code is driven by biotic factors (the ``Red Queen'' model) or abiotic factors (the ``Court Jester'' model), a topic which is itself an area of debate in the biological sciences~\cite{Benton2009}.  On the one hand, our code mutation proposals can be considered biotic factors and can introduce short-term ecosystem changes. On the other hand, the longer term incentives from the dominant business models of the era may abiotically dominate any such localised strategies.  More systemic changes may therefore be required to make room for evolutionary approaches.

\section{Conclusions}
\label{s:conclusions}
The Internet connects around ten billion people and systems. Within your biome alone, there are more than ten trillion microbial cells. And there are 10 million different species of creatures. This scale and diversity survives a wide range of environments and environmental change and, while messy, deserves attention for the wide range of mechanisms that serve that survival, as an inspiration for future technologies. 
The lack of gene-pool diversity in the Internet and its associated risks has been commented on for over twenty years.\footnote{\url{https://web.archive.org/web/20030425082402/http://www.icir.org/mjh/nyt-orig.html}} At around the same time, the potential of network systems to evolve was observed by Doyle: {\it ``TCP/IP is an example of how architectures that are well-designed for extreme robustness can create evolvability as a side benefit}''~\cite{doi:10.1073/pnas.1103557108}. Today, we still retain a fundamental optimism towards the long term health and ability of the Internet to resist capture into a singular monoculture.

However, we are not alone in recognising the need for a shakeup \textit{now} before the ossification sets in too deeply to correct. Farrell and Berjon also recently observed that the ``\textit{drive to centralize, control and extract}'' has driven the Internet towards an ossified pathology of command and control\footnote{\url{https://www.noemamag.com/we-need-to-rewild-the-internet/}} and made a powerful case for ``rewilding'' the Internet.
We should collectively derive inspiration from the principles of governing the commons from Ostrom~\cite{Ostrom2015}, where one main conclusion already sees that rule-based protocols will not scale, and suggest the need to move towards nested ecologically-driven approaches and incentives.
We therefore believe it is now the right time to embrace these messy but extraordinarily robust and sustainable mechanisms from nature into artificial systems such as the Internet, so that they too can survive the current crisis and grow and thrive into the coming century.

\section*{Acknowledgements}
We gratefully acknowledge our many colleagues in the Cambridge Conservation Initiative and the University of Cambridge Energy and Environment Group who helped debate and shape the arguments in this paper. We thank AI@CAM and the UKRI CRCRM TFS (MR/Y024354) for funding portions of this work, as well as unrestricted donations from Tarides, John Bernstein and Jane Street.
Some authors received financial support from Imperial College London through an Imperial College Research Fellowship grant, as well as a Henslow Fellowship funded by the Cambridge Philosophical Society.  The funders had no role in the design, analysis, decision to publish, or preparation of the manuscript.

\printbibliography

\end{document}